\documentclass[twocolumn,aps,amsfonts,nofootinbib,showpacs%,tightenlines
]{revtex4}

\usepackage{epsf}
\usepackage{amscd}
\usepackage{amsmath}
\usepackage{graphicx,comment}
\input xy
\xyoption{all}

\def\ol#1{{\overline{#1}}}

\newcommand{\beq}{\begin{equation}}
\newcommand{\eeq}{\end{equation}}
\newcommand{\bea}{\begin{eqnarray}}
\newcommand{\eea}{\end{eqnarray}}

\newcommand{\nn}{\nonumber}
\newcommand{\benn}{\begin{displaymath}}
\newcommand{\eenn}{\end{displaymath}}
\newcommand{\x}{{\mathbf{x}}}

\newcommand{\e}{\mathbf e}

\def\slashchar#1{\ensuremath{                               %
   \setbox0=\hbox{${}#1{}$}       % set a box for #1 
   \dimen0=\wd0                                 % and get its size
   \setbox1=\hbox{/} \dimen1=\wd1               % get size of /
   \ifdim\dimen0>\dimen1                        % #1 is bigger
      \rlap{\hbox to \dimen0{\hfil/\hfil}}      % so center / in box
      {}#1{}                                    % and print #1
   \else                                        % / is bigger
      \rlap{\hbox to \dimen1{\hfil${}#1{}$\hfil}}   % so center #1
      /                                         % and print /
   \fi}}                                        %

\begin{document}

\preprint{UMD-40762-41x}

\author{P.~F.~Bedaque}
\email[]{bedaque@umd.edu}
\affiliation{%
Maryland Center for Fundamental Physics, 
Department of Physics, 
University of Maryland, 
College Park,  
MD 20742-4111, 
USA
}

\author{M.~I.~Buchoff}
\email[]{mbuchoff@umd.edu}
\affiliation{%
Maryland Center for Fundamental Physics, 
Department of Physics, 
University of Maryland, 
College Park,  
MD 20742-4111, 
USA
}

\author{B.~C.~Tiburzi}
\email[]{bctiburz@umd.edu}
\affiliation{%
Maryland Center for Fundamental Physics, 
Department of Physics, 
University of Maryland, 
College Park,  
MD 20742-4111, 
USA
}

\author{A.~Walker-Loud}
\email[]{walkloud@umd.edu}
\affiliation{%
Maryland Center for Fundamental Physics, 
Department of Physics, 
University of Maryland, 
College Park,  
MD 20742-4111, 
USA
}

\title{Search for Fermion Actions on Hyperdiamond Lattices}

\begin{abstract}
Fermions moving in a two-dimensional honeycomb lattice (graphene) have, at low energies, chiral symmetry.  Generalizing this  construction to four dimensions potentially provides fermions with chiral symmetry and only the minimal fermion doubling demanded by the Nielsen-Ninomiya no-go theorem. The practical usefulness of such fermions hinges on whether the action has a necessary set of discrete symmetries of the lattice.  If this is the case, one avoids the generation of dimension three and four operators which require fine tuning.  We construct hyperdiamond lattice actions with enough symmetries to exclude fine tuning; however, they produce multiple doublings.  The limit where the actions exhibit minimal doubling does not possess the requisite symmetry.
\end{abstract}


\maketitle

\section{Introduction}
When naive, chirally symmetric discretizations of spin one-half fermions are used, 
the familiar phenomenon of doubling arises: 
even though only one Dirac fermion is attached to each lattice node, the continuum limit describes several fermions. 
This is a problem for most applications, like QCD, where only a small number of light quarks exist. For the naive fermion action in four space-time dimensions, doubling leads to $2^{4}=16$ fermions. More complicated discretizations can reduce the number of doublings but not eliminate them completely, at least not while preserving exact chiral invariance. This is essentially the content of the celebrated Nielsen-Ninomiya ``no-go" theorem~\cite{nielsen}. 

In the past, Karsten~\cite{karsten} and Wilczek~\cite{wilczek} proposed fermion actions with good chiral symmetry and the minimal fermion doubling allowed by the no-go theorem. These actions were not successful in practice because they fail to have the correct continuum limit. The cuplrit of the problem is the lack of enough symmetry to forbid the existence of dimension three and four operators in the low-energy effective theory describing the lattice artifacts. Thus, these actions require additional terms with coefficients fine tuned in order for the correct continuum limit to arise as the lattice spacing goes to zero. Usually fine tunings of this kind are very difficult to implement.

Recently, Creutz~\cite{creutz} and, shortly after, Bori\c{c}i~\cite{borici} suggested a 
two-parameter class of actions with minimal doubling that generalized graphene to four space-time dimensions. Their action can be interpreted as being defined on a four-dimensional generalization of the two-dimensional honeycomb lattice (which is sometimes referred to as a hyperdiamond lattice). It was subsequently pointed 
out~\cite{bedaque} that this action lacks enough symmetry to prevent dimension three and four operators from being generated. The hyperdiamond lattice, however, has a symmetry under the permutation of any two out of five distinct axes. This $S_5$ symmetry (or a large enough subgroup, such as the group of cyclic permutations $\mathbb{Z}_5$), if possessed also by the action, would eliminate all dimension three and four operators that cause the fine tuning problem. A natural question then is whether the $\mathbb{Z}_5$ symmetry can be recovered for special values of the action parameters, as suggested in~\cite{creutz}, or whether, failing that, a modified $\mathbb{Z}_5$ symmetric action can be engineered. In order to clarify this issue, we construct  
a $\mathbb{Z}_5$ symmetric action on a hyperdiamond lattice. 
This action exhibits multiple doublings, and attempts to eliminate these doublings 
lead to other undesirable features.

\section{Hyperdiamond lattice action}
\subsection{Hyperdiamond lattice}
We start the construction of the four-dimensional hyperdiamond lattice by defining a set of five distinct vectors $\{ \e^\alpha \}$ satisfying
\bea\label{eq:e_properties}
\e^\alpha \cdot \e^\beta &=& \left\{ \begin{matrix}
                               \phantom{--}1 ,\ \  {\rm for} \ \  \alpha=\beta\\
                             \cos \theta,\ \  {\rm for} \ \  \alpha\neq\beta
                        \end{matrix} \right.
.\eea 
Straightforward linear algebra shows that 
the mutual angle between these vectors
is given by
$\cos \theta = - 1 / 4$. 
Consequently we have the relation
%\bea
$\sum_{\alpha=1}^5 \e^\alpha =0$. 
%\eea
%
%
We can take the vectors $\e^\alpha$ to be
\begin{align}
\label{eq:es}
\e^1 &= \frac{1}{4} (\phantom{+}\sqrt{5},\phantom{+}\sqrt{5},\phantom{+}\sqrt{5},\ 1),\nn\\
\e^2 &= \frac{1}{4} (\phantom{+}\sqrt{5},-\sqrt{5},-\sqrt{5},\ 1),\nn\\
\e^3 &= \frac{1}{4} (-\sqrt{5},-\sqrt{5},\phantom{+}\sqrt{5},\ 1),\nn\\
\e^4 &= \frac{1}{4} (-\sqrt{5},\phantom{+}\sqrt{5},-\sqrt{5},\ 1),\nn\\
\e^5 &= -  ( \phantom{++}0,\phantom{++}0,\phantom{++}0,\,\ 1) .
\end{align} 
The first four elements of this set form a basis whose dual basis $\{ \e_\mu \}$ is given, 
up to normalization, by
\bea\label{eq:e_dual}
\e_1 &=& \e^1- \e^5,\nn\\
\e_2 &=& \e^2- \e^5,\nn\\
\e_3 &=& \e^3- \e^5,\nn\\
\e_4 &=& \e^4- \e^5.
\eea 
The hyperdiamond lattice is formed by two sublattices. One (the ``L-nodes") are generated by integer combinations of
$\e_\mu$, 
that is, the points 
$\x_L=\sum_{\mu=1}^4 x^\mu \e_\mu$, 
with integer 
$x^\mu$. 
The second sublattice (the ``R-nodes") is obtained from the first by translating it by 
$\e^5$, 
so that it is formed by the points 
$\x_R=\sum_{\mu=1}^4 x^\mu \e_\mu+\e^5$.
Notice that the four coordinates 
$x^\mu$ do not specify a node uniquely: 
we need to specify, in addition, whether the point is an L-node or an R-node.

The five nearest neighbors of an L-node with coordinates 
$x^\mu$ 
are the R-nodes with coordinates 
\begin{eqnarray}
&& (x^1+1, x^2, x^3, x^4), \quad
(x^1, x^2+1, x^3, x^4), \notag \\ 
&& (x^1, x^2, x^3+1, x^4),  
\quad 
(x^1, x^2, x^3, x^4+1), \quad \text{and}  \notag \\ 
&&  (x^1, x^2, x^3, x^4), \notag 
\end{eqnarray}
as is easily verified. Similarly, the nearest neighbors of an R-node with coordinates 
$(x^1, x^2, x^3, x^4)$ 
are the L-nodes with coordinates
\begin{eqnarray}
&& (x^1-1, x^2, x^3, x^4), \quad
(x^1, x^2-1, x^3, x^4), \notag \\ 
&& (x^1, x^2, x^3-1, x^4),  
\quad 
(x^1, x^2, x^3, x^4-1), \quad \text{and}  \notag \\ 
&&  (x^1, x^2, x^3, x^4) \notag 
.\end{eqnarray}
The separation between one node and its nearest neighbors are the five vectors 
$\e^\alpha$.
The symmetric distribution of these vectors 
means that the nearest neighbors are also distributed symmetrically around each node. 
In fact, not only the nearest neighbors but the whole hyperdiamond lattice is symmetrical 
under any permutation of the five 
$\e^\alpha$ 
axes. 
Hence the hyperdiamond lattice has an $S_5$ symmetry.
These permutations are of crucial importance, and so we examine them carefully.  
The permutations involving only the first four axes are described in the coordinate system 
used here by the same permutation of the 
$x^\mu$ 
coordinates. For instance, exchange of the 
$\e^1$ 
and 
$\e^2$ 
axes is the mapping
\begin{equation} \label{eq:transf}
%x = 
(x^1, x^2, x^3, x^4) \longrightarrow \tilde{x} = (x^2, x^1, x^3, x^4)
,\end{equation}
for both L-nodes and R-nodes. 
An exchange involving the fifth axis is slightly more complex. 
The
$\e^1$
and
$\e^5$ 
exchange, for instance, takes the L-node
\begin{equation}\label{eq:Ltransf}
%x = 
(x^1, x^2, x^3, x^4)
\longrightarrow
\tilde{x} =  \big(-(x^1+ x^2+ x^3+ x^4), x^2, x^3, x^4\big)
,\end{equation} 
but takes the R-node 
\begin{equation} \label{eq:Rtransf}
%x = 
(x^1, x^2, x^3, x^4)
\longrightarrow 
\tilde{x} = \big(-(x^1+ x^2+ x^3+ x^4)+1, x^2, x^3, x^4\big)
.\end{equation}

\subsection{Hyperdiamond action}

We can build a free fermion action on the hyperdiamond lattice by attaching 
a two-component left-handed spinor $\phi$ 
and right-handed spinor $\ol\phi$ 
to each L-node, 
and a right-handed spinor $\chi$ and left-handed spinor $\ol\chi$ to every R-node. 
The action
\bea\label{eq:Z5}
S &=&  \sum_x \Bigg[\sum\limits_{\mu=1}^4\left(  \,
 \ol\phi_{x-\mu} \, \sigma \cdot \e^\mu \, \chi_{x}
 -\ol\chi_{x+\mu} \, \ol\sigma \cdot \e^\mu \, \phi_x \right)
 \notag \\
&& \phantom{spacing}
 +\ol\phi_{x} \, \sigma \cdot \e^5 \, \chi_{x} 
 -\ol\chi_{x} \, \ol\sigma \cdot \e^5 \, \phi_x\Bigg],
\eea 
with 
$\ol\sigma=(\vec{\sigma}, -i)$, 
$\sigma=(\vec{\sigma}, i)$, 
describes fermions hopping to nearest-neighbor sites with equal probability in all five directions. Since $\phi_x$ and $\ol\chi_x$ live on L- and R-nodes, respectively, the last two terms describe the hopping along the $\e^5$ direction even though the coordinate $x$ is the same for both fields.% 
\footnote{
In momentum space, this action reads
\beq
S=
%\frac{1}{2}
\int_p 
\ol\psi_p \left[ 
	i\sum_\mu \sin(p_\mu) \e^\mu \cdot \gamma
	-\Big(\sum_\mu \cos(p_\mu) \e^\mu+\e^5\Big) \cdot \gamma\, \gamma_5
\right] \psi_p\nn
,\eeq with
\beq
\psi_p = \begin{pmatrix}
\phi_p\\
\chi_p
\end{pmatrix},
\quad 
\ol \psi_p  
= 
\left( \ol \phi_p , \ol \chi_p  \right),
\quad 
\text{and}
\quad 
\gamma_\mu
=
\begin{pmatrix}
0 & \sigma_\mu\\
 \ol\sigma_\mu  & 0 
 \end{pmatrix}
.\nn
\eeq 
}%%END OF FOOTNOTE

The action in Eq.~(\ref{eq:Z5}) is not invariant under the full 
$S_5$ symmetry group of the lattice.  However, it is invariant under the subgroup of transformations which can be built by an even number of elementary permutations of two axes, known as the alternating group, $A_5$.  Under an $A_5$ transformation, the fields transform as
\begin{align}
\label{eq:permutation}
&\phi_x {\longrightarrow} \, P_{\alpha\beta} \,  \ol P_{\gamma\delta} \,  \phi_{\tilde{\tilde{x}}},\ \ \ \ 
&\ol\chi_x {\longrightarrow} \, \ol\chi_{\tilde{\tilde{x}}} \, \ol P_{\gamma\delta} \, P_{\alpha\beta} ,\nn\\
&\chi_x {\longrightarrow} \, \ol P_{\alpha\beta} \, P_{\gamma\delta}\, \chi_{\tilde{\tilde{x}}},\ \ \ \ 
&\ol\phi_x {\longrightarrow} \, \ol\phi_{\tilde{\tilde{x}}} \, P_{\gamma\delta}\, \ol P_{\alpha\beta} \, ,
\end{align} 
where the permutation operators are 
\beq
P_{\alpha\beta} = i\frac{\e^\alpha-\e^\beta}{|\e^\alpha-\e^\beta|}\cdot\sigma \, ,\ \ \
\ol P_{\alpha\beta} = i\frac{\e^\alpha-\e^\beta}{|\e^\alpha-\e^\beta|}\cdot\ol\sigma \, ,
\eeq 
and the coordinates transform as in Eqs.\eqref{eq:transf}--\eqref{eq:Rtransf}.  For example, under a transformation involving a permutation of the axes $\e^1$ and $\e^2$,
as well as a permutation of $\e^1$ and $\e^3$, the field
$\phi_x$ transforms as
\begin{equation} %\notag
\phi_x \longrightarrow P_{12} \, \ol P_{13} \, \phi_{x_3, x_1, x_2, x_4}
.\end{equation}
The transformation of the action in Eq.~(\ref{eq:Z5}) under permutations of the five axes follows from
\bea
P_{\alpha\beta} \, \ol\sigma \cdot \e^\alpha  \, P_{\alpha\beta} &=& \sigma \cdot \e^\beta, \nn\\
\ol P_{\alpha\beta} \, \sigma \cdot \e^\alpha  \, \ol P_{\alpha\beta} &=& \ol \sigma \cdot \e^\beta, \nn\\
P_{\alpha\beta} \, \ol\sigma \cdot \e^\gamma \, P_{\alpha\beta} &=& \sigma \cdot \e^\gamma,\ \alpha,\beta\neq\gamma \nn\\
\ol P_{\alpha\beta} \, \sigma \cdot \e^\gamma \, \ol  P_{\alpha\beta} &=& \ol \sigma \cdot \e^\gamma,\ \alpha,\beta\neq\gamma  
\eea 
which, in turn, follows from
\beq
2 a\cdot b \,  b \cdot \sigma =(a \cdot \sigma  b \cdot \ol\sigma + b \cdot \sigma a \cdot \ol\sigma) b \cdot \sigma
= a \cdot \sigma + b \cdot \sigma \, a \cdot \ol\sigma \, b \cdot \sigma,
\eeq 
valid for any unit vectors $a$ and $b$.  Being careful to remember that L-nodes and R-nodes transform differently, and that $\phi$, $\ol \phi$ sit on L-nodes, while $\chi$, $\ol \chi$ sit on R-nodes, one can verify that the action, Eq.~\eqref{eq:Z5}, is invariant under the subgroup of even permutations of $S_5$, the alternating group, $A_5$.

There are only two chirally symmetric relevant operators in the continuum
that are $A_5$ symmetric:
$i \sum_\alpha \ol \psi(x)   \, \e^\alpha \cdot \gamma \, \psi(x)$,
and
$\sum_\alpha \ol \psi(x)  \, \e^\alpha \cdot \gamma \, \gamma_5 \, \psi(x)$.
Both of these operators, however, identically vanish because the five
basis vectors sum to zero.
In fact, we need not require invariance under the full $A_5$ symmetry
to exclude such operators, the cyclic permutation subgroup $\mathbb{Z}_5 \subset A_5$
is the minimal subgroup of $S_5$ that excludes relevant operators. 
While the action exhibits enough symmetry for a good continuum limit, it does not, however, have minimal doubling. 
In fact, it has, in addition to the pole at $p_\mu=0$, where the action reduces to the Dirac form, several others poles at finite $p_\mu$, \textit{e.g.} $p_1 = -p_2 = -p_3 = p_4 = \cos^{-1}(-2/3)$. 
Let us now consider actions with minimal doubling and argue that, unfortunately, they seem to lack the crucial $\mathbb{Z}_5$ symmetry.

\section{Comparison with other actions}
Creutz \cite{creutz}, and later Bori\c ci \cite{borici}, suggested minimally doubled 
fermion actions based on similar considerations to the ones discussed above. 
In fact, the Bori\c ci-Creutz action can be written in the form
\begin{eqnarray} 
S_{BC} 
&=& 
\frac{1}{2}\sum_{x}
\Bigg[
\sum_\mu
\left(\ol \psi_{x- \mu}
\, \e^\mu \cdot \Gamma \,
\psi_{x}
-
\ol \psi_{x+\mu}
\,
\e^\mu \cdot \Gamma^\dagger
\,
\psi_{x}
\right)
\notag \\
&& \phantom{spacing}+
\ol \psi_x
\,
\e^5 \cdot \Gamma
\,
\psi_x-\ol \psi_x
\,
\e^5 \cdot \Gamma^\dagger
\,
\psi_x
\Bigg],
\label{eq:CreutzZ5}
\end{eqnarray}
where 
$\Gamma_\mu = (\vec{\gamma}, i \gamma_4)$ and the vectors $\e^\alpha$ are defined
in terms of two parameters $B$ and $C$ as
\bea\label{eq:es2}
\e^1 &=& \phantom{-}(\phantom{+}1,\phantom{+}1,\phantom{+}1,\phantom{4}B\ ),\nn\\
\e^2 &=& \phantom{-}(\phantom{+}1,-1,-1,\phantom{4}B\ ),\nn\\
\e^3 &=& \phantom{-}(-1,-1,\phantom{+}1,\phantom{4}B\ ),\nn\\
\e^4 &=& \phantom{-}(-1,\phantom{+}1,-1,\phantom{4}B\ ),\nn\\
\e^5 &=& -(\phantom{+}0,\phantom{+}0,\phantom{+}0,4BC).
\eea 
%This action is similar to Eq.~(\ref{eq:Z5}) and, 
In order to compare this action to
Eq.~(\ref{eq:Z5}), 
let us write it in terms of the left- and right-handed components 
of the Dirac spinor $\psi$:%
\bea\label{eq:BC_twocomponent}
S_{BC} 
&=& 
\frac{1}{2}\sum\limits_x 
\Bigg[
\sum\limits_{\mu}\left(
\ol\phi_{x-\mu} \, \Sigma \cdot \e^\mu \, \chi_{x}
 -\ol\chi_{x+\mu} \, \Sigma \cdot \e^\mu \, \phi_x  \right)
 \notag \\
 && 
 \phantom{spacinging}
 +\ol\phi_{x} \, \Sigma \cdot \e^5 \, \chi_{x} 
 -\ol\chi_{x} \, \Sigma \cdot \e^5 \, \phi_x
  \nn\\
&& \phantom{spaci} + \sum\limits_\mu
 \left(\ol\chi_{x-\mu} \, \ol\Sigma \cdot \e^\mu \, \phi_{x}
 -\ol\phi_{x+\mu} \, \ol\Sigma \cdot \e^\mu \, \phi_x \right)
 \notag \\
 && \phantom{spacinging}
  +\ol\chi_{x} \, \ol\Sigma \cdot \e^5 \, \phi_x
-\ol\phi_{x} \, \ol\Sigma \cdot \e^5 \, \chi_{x} 
\Bigg],
\eea with $\Sigma=(\vec{\sigma}, -1)$ and $\ol\Sigma=(\vec{\sigma}, 1)$. 

\begin{figure}[h]
\center
\begin{tabular}{c}
\includegraphics[width=0.8\columnwidth]{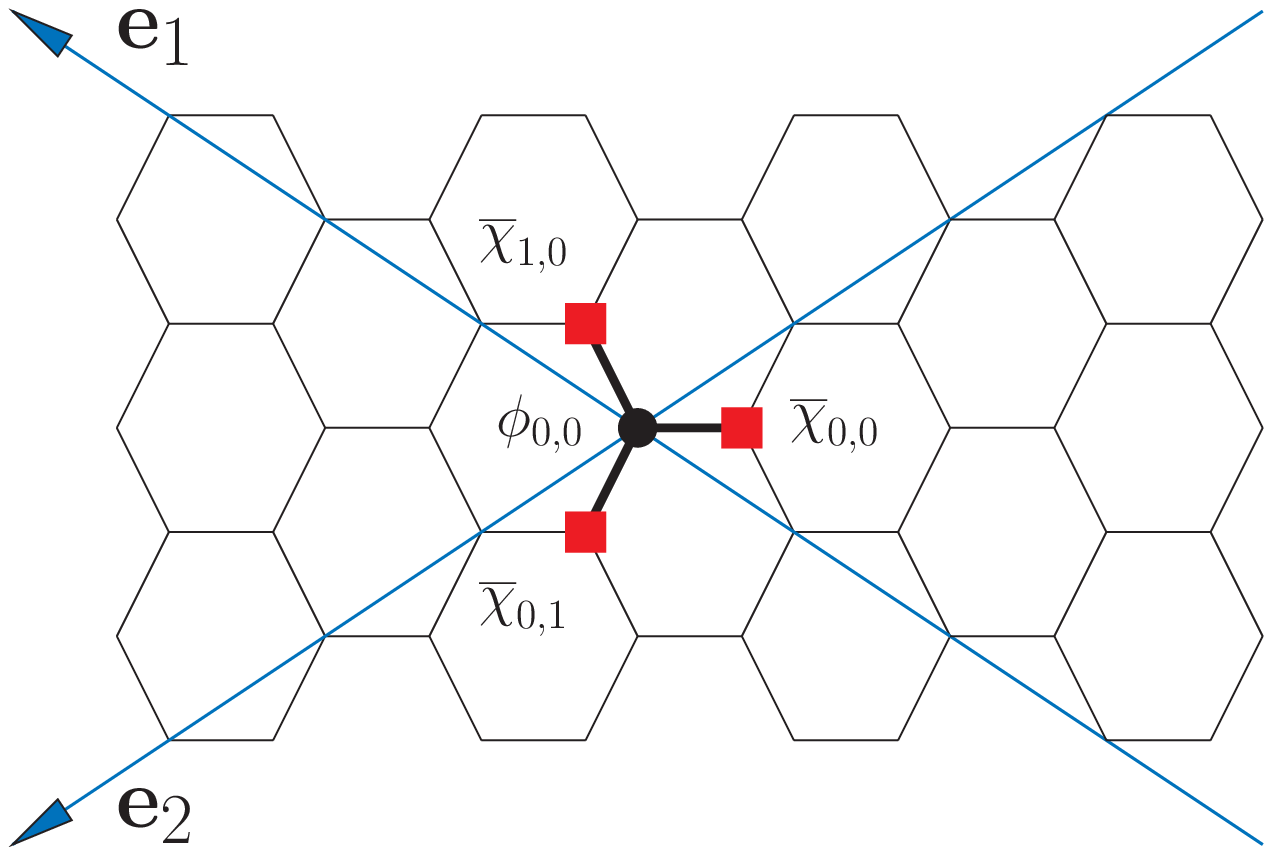}
\\ \\
\includegraphics[width=0.8\columnwidth]{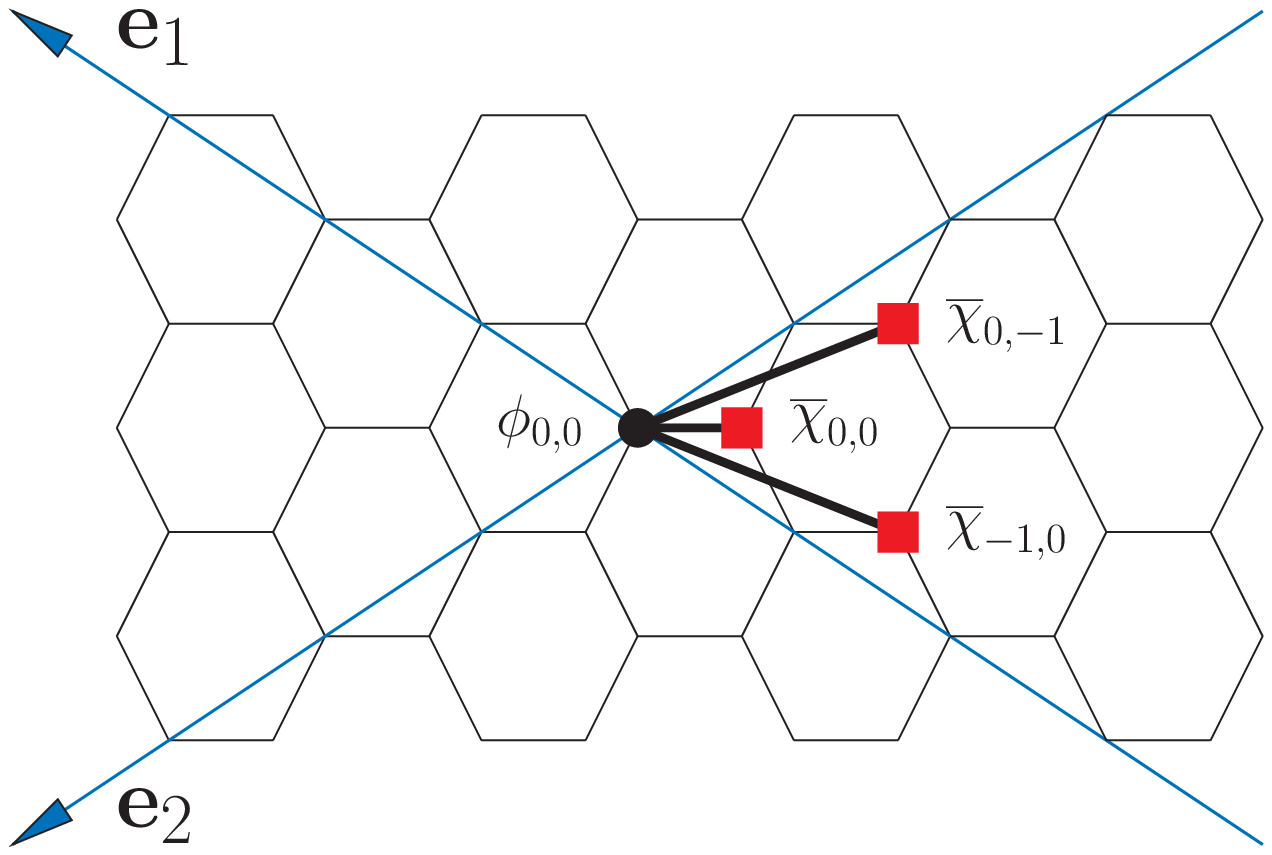}
\end{tabular}
\caption{\label{fig:Good_Bad} The two-dimensional analogue of the hyperdiamond lattice.  The long (blue) lines denote the dual basis, $\{\mathbf{e}_1,\mathbf{e}_2\}$, as defined in Eq.~\eqref{eq:e_dual}.  In both figures, the (black) circle sits on an L-node, and denots the field $\phi_{0,0}$.  The (red) squares denote various $\ol{\chi}$ fields sitting on R-nodes.  \textit{Top figure:} the three nearest neighbor fields, $\ol{\chi}_{0,0}$, $\ol{\chi}_{1,0}$ and $\ol{\chi}_{0,1}$, connected by thick (black) links.  \textit{Lower figure:} the three (non)-nearest neighbor fields, $\ol{\chi}_{0,0}$, $\ol{\chi}_{-1,0}$ and $\ol{\chi}_{0,-1}$, connected by thick (black) links, present in the $S_{BC}$ action.}
\end{figure}

Despite the similarity between $S$ in Eq.~(\ref{eq:Z5}) and $S_{BC}$ in Eq.~(\ref{eq:BC_twocomponent}), there are important differences. First of all, 
 for generic $B$ and $C$, the vectors $\e^\alpha$ are not symmetrically arranged and thus  break the $\mathbb{Z}_5$ symmetry of the lattice.
Only the case $B=1/\sqrt{5}$ and $C=1$ satisfies Eq.~\eqref{eq:e_properties},
up to the overall normalization.
A further difference is that
there is an extra factor of $i$ in the fourth components of $\Sigma$, $\ol \Sigma$ compared to $\sigma$, $\ol \sigma$. The reason both actions are able to reproduce the Dirac propagator despite this additional $i$ factor is that while $S$ generates a pole at $p_\mu=0$, the poles of $S_{BC}$ are at finite $p_\mu$, where it duly reduces to the Dirac action. The different pole positions also explain why the position of $\Sigma$, $\ol \Sigma$  appear shifted from the positions of  $\sigma$, $\ol\sigma$, even though both actions reduce to the Dirac action around their poles.  
The differences in spin structure 
do not spoil the spinor part of the $S_5$ transformations
provided that $B = 1 / \sqrt{5}$, and $C=1$.
Finally,
$S_{BC}$ contains twice as many terms as $S$:
the additional terms comprise the last two lines of Eq.~(\ref{eq:BC_twocomponent}). 
These extra terms can be visualized as 
non-nearest neighbor interactions in the hyperdiamond lattice. 
To see this, take the case where $x=(0,0,0,0)$. 
As shown in Fig.~\ref{fig:Good_Bad}, the term $\ol\chi_{x-1} \, \ol\Sigma \cdot \e^1 \, \phi_{x}$, for instance, describes a hopping from the point $\bf{x}=\bf{0}$ to $-\e_1+\e^5$. Most importantly, the additional terms in the last two lines of Eq.~\eqref{eq:BC_twocomponent} 
break the $\mathbb{Z}_5$ symmetry under coordinate transformations.  
Without the additional symmetry, relevant and marginal operators
will be generated by gauge interactions~\cite{bedaque}.

%
%\begin{widetext}
%
%%%%%%%%%     FIGURE   Good Bad  %%%%%%%%%%%
%\end{widetext}

\section{Modified actions}
Guided by our construction of an $A_5$ symmetric action on the hyperdiamond lattice,
we can try to answer the question:
is there a way of combining the good features of the Bori\c ci-Creutz action (minimal doubling and chiral symmetry) while avoiding its pitfall (lack of $\mathbb{Z}_5$ symmetry)? 
We have been unable to find such an action which answers this question. 
For instance, one possibility is to 
demand invariance under $\mathbb{Z}_5$
and hence drop the terms in the last two lines of Eq.~\eqref{eq:BC_twocomponent}. 
This corresponds to including chiral projection operators in Eq.~\eqref{eq:CreutzZ5}.
We must further choose $B=1 / \sqrt{5}$ and $C=1$.
The resulting action is $A_5$ symmetric 
with a pole at $p_\mu=0$, and several other values of $p_\mu$. 
The pole at $p_\mu = 0$, due to the misplaced $i$ factors and $\Sigma$ matrices
relative to Eq.~\eqref{eq:Z5}, 
has the wrong form. The action around $p_\mu\simeq 0$ behaves like
$\ol\psi_k (i\vec{\gamma} \cdot \vec{k} +  \gamma_5\gamma_4 k_4)\psi_k$,
for suitably defined physical particle momenta $k_\mu = k_\mu(p_\nu)$.
This form fails to reproduce the Dirac equation in the naive continuum limit. 
Fermion poles of this form, sometimes called mutilated fermions, 
also appear in other attempts at formulating fermions on non-hypercubic 
lattices~\cite{celmaster,drouffe}.

\section{Conclusion}

By analogy with fermions on a honeycomb lattice, 
we construct a hyperdiamond fermion lattice action. 
This action possesses a large subgroup ($A_5$) of the  $S_5$ symmetry of 
the hyperdiamond lattice, and 
consequently enough symmetry to avoid
fine-tunings in taking the continuum limit.
The hyperdiamond action has, however, 
more than the minimal amount of fermion doubling.

We investigate the lattice actions proposed 
by Creutz and Bori\c ci, and find that, while 
similar to the hyperdiamond construction,
these actions do not have at least a $\mathbb{Z}_5$ symmetry.
Hence the continuum limit of these theories 
will require fine tuning.
Modification of these actions to enforce
$\mathbb{Z}_5$ symmetry from the outset leads to undesirable 
effects such as additional poles,
and a Lorentz non-symmetric continuum limit.

The goal of our search is to produce 
a minimally doubled action with 
a good continuum limit.
We find, however, 
an intricate balance needed to break hypercubic, 
parity, and time-reversal symmetry in order to 
obtain minimal doubling, while, at the same
time, preserve or invent additional symmetries 
necessary to avoid fine tunings.
The requirement of $\mathbb{Z}_5$ symmetry 
on a hyperdiamond appears incommensurate 
with minimal doubling.

\begin{acknowledgments}
We thank M.~Golterman and Y.~Shamir for correspondence.
This work is supported in part by the 
U.S.~Dept.~of Energy,
Grant No.~DE-FG02-93ER-40762-410.
\end{acknowledgments}

%%%%%%%%%%%%%%%%%%%%%%%%%%%%%%%%%%
  
\end{document}